%% file: main.tex
\documentclass{article}
\usepackage{spconf}
\usepackage{amsmath,amsfonts,amssymb,amsthm}
\usepackage{graphicx}
\usepackage{float}
\usepackage{subcaption}
\usepackage{algorithm,algpseudocode} 
\usepackage{bm}
\usepackage{xcolor}
\usepackage{multirow}

\title{KELP: K-space-conditioned Estimation of Learned Sampling Patterns for Scan-Adaptive Multi-Coil MRI}

\name{Aryan Dhar$^{1,*}$ \qquad Siddhant Gautam$^{2,*}$ \qquad Saiprasad~Ravishankar$^{1,3}$
\thanks{$^{*}$A. Dhar and S. Gautam contributed equally to this work. S. Gautam (\textit{gautamsi@umich.edu}) and S. Ravishankar (\textit{ravisha3@msu.edu}) are the corresponding authors. This work was partially supported by {National Institutes of Health} Grant Number {R21 EB030762.}}}

\address{$^{1}$ Computational Mathematics Science and Engineering, Michigan State University, East Lansing, MI, USA \\
$^{2}$ Electrical and Computer Engineering, University of Michigan, Ann Arbor, MI, USA\\
$^{3}$ Biomedical Engineering, Michigan State University, East Lansing, MI, USA}

\begin{document}
\ninept

\maketitle

\begin{abstract}
Deep learning techniques have gained considerable attention for accelerating MRI acquisition while maintaining image quality. In this work, we present a convolutional neural network (CNN)-based framework for predicting scan-adaptive undersampling patterns directly from low-frequency multi-coil $k$-space data. Unlike approaches that optimize sampling patterns during training or rely on nearest-neighbor search at inference, our method is trained using precomputed scan-adaptive optimized masks as supervised labels and predicts a scan-specific sampling pattern in a single forward pass. The training procedure alternates between optimizing the sampling network and a reconstruction network, allowing the learned masks to adapt to reconstruction performance. 
Experiments on the fastMRI multi-coil knee dataset demonstrate competitive reconstruction quality compared with existing population- and scan-adaptive sampling approaches at $4\times$ and $8\times$ acceleration factors. In addition, the proposed method enables efficient scan-specific mask prediction with inference cost independent of the training dictionary size.
\end{abstract}


\section{Introduction}
\input{writeup/introduction}

\section{Methodology}
\input{writeup/methodology}

\section{Experiments $\&$ Results}
\input{writeup/experiments}

\section{Conclusion}
\input{writeup/conclusion}


\bibliographystyle{IEEEbib}
\bibliography{references}

\end{document}

%% file: writeup/introduction.tex
Magnetic Resonance Imaging (MRI) is an important tool in modern medical diagnostics, offering non-invasive, high-quality visualization of soft tissues without ionizing radiation. However, MRI acquisition is inherently time-consuming, which can increase patient discomfort, reduce scanner throughput, and limit accessibility. Accelerating data acquisition while preserving diagnostic image quality has therefore remained a central problem in MRI.
One way to accelerate acquisition is to undersample the phase-encoding direction, although doing so introduces aliasing in the reconstructed image. Parallel imaging (PI) addresses this by exploiting the distinct spatial sensitivity profiles of multiple receiver coils to provide additional encoding information and recover images from undersampled data~\cite{pruessmann2006encoding, ying2010parallel}. Compressed sensing (CS) provides an alternate acceleration technique by exploiting signal structure such as sparsity together with incoherent sampling~\cite{donoho2006compressed, lustig2007sparse}. In both settings, reconstruction quality depends strongly on which $k$-space measurements are acquired, making the design of effective undersampling patterns a key component of accelerated MRI.

Early data-driven approaches optimized a single sampling pattern over a training population using greedy or subset-selection techniques~\cite{gozcu2018learning, gozcu2019rethinking, sanchez2020scalable, zibetti2021fast, zibetti2022alternating}. More recently, deep learning has enabled joint optimization of sampling and reconstruction networks~\cite{bahadir2020deep, zhang2020extending, aggarwal2020j, sherry2020learning, yin2021end, huang2022single, alkan2024autosamp}. 
For example, LOUPE~\cite{bahadir2020deep} jointly learns a probabilistic $k$-space sampling mask and a reconstruction network under a prescribed sampling budget, with the learned probabilities converted into a realizable binary sampling pattern. J-MoDL~\cite{aggarwal2020j} similarly optimizes sampling parameters jointly with an unrolled MoDL reconstructor. These methods can produce effective population-adaptive sampling patterns, but typically apply the same learned sampling pattern across scans.

Another class of methods performs scan-adaptive sampling, in which the initial $k$-space measurements from an individual scan are used to determine which additional $k$-space locations should be sampled~\cite{yin2021end, huang2022single, gautam2024patient, gautam2026scan, gautam2026dsuno}. Such adaptation can exploit scan-specific anatomical and spectral information that is not captured by a single population-level pattern. The central challenge is to obtain high-quality scan-specific sampling masks while keeping inference sufficiently fast and simple for prospective acquisition.
MNet~\cite{huang2022single} predicts a scan-specific sampling pattern directly from low-frequency $k$-space using a convolutional neural network (CNN). SUNO~\cite{gautam2024patient, gautam2026scan, gautam2026dsuno}, in contrast, generates high-quality scan-specific masks offline on a training set (dictionary) using iterative coordinate descent (ICD)-based sampling optimization and then estimates an optimized mask using nearest-neighbor matching of the low-frequency $k$-space at inference. 

In this work, we combine the advantages of MNet-based test-time mask prediction and the scan-adaptive nature of ICD masks (SUNO) for training through {pre-optimized mask supervision}. Given the low-frequency $k$-space measurements of a new scan, the trained network directly predicts its undersampling pattern without performing iterative optimization or searching through a database of training examples. We develop this framework for multi-coil MRI and jointly optimize the sampler and reconstruction network through alternating updates. Experiments on the fastMRI multi-coil knee dataset demonstrate improved reconstruction quality compared with existing population- and scan-adaptive sampling approaches across both 4$\times$ and 8$\times$ acceleration factors. In addition, the proposed method enables efficient scan-specific mask prediction with inference cost independent of the training dictionary size.

%% file: writeup/methodology.tex
\subsection{Multi-coil MRI Reconstruction Framework}
In multi-coil MRI, multiple receiver coils acquire $k$-space data simultaneously, each with a distinct spatial sensitivity profile. The goal of the multi-coil MRI reconstruction is to recover the underlying image $\mathbf{x}$ from these undersampled multi-coil $k$-space measurements. The signal from the $c$-th coil can be modeled as the product of the image and the coil's sensitivity map $\mathbf{S}_c$, followed by a Fourier transform $\mathbf{F}$ and a sampling mask $\mathbf{M}$ that selects the acquired $k$-space locations. The measured $k$-space data $\mathbf{y}_c$ for $C$ coils can be expressed as:
\begin{equation}
    \mathbf{y}_c = \mathbf{M}\mathbf{F}\mathbf{S}_c \mathbf{x} + \mathbf{n}_c, \quad c = 1, \ldots, C,
\end{equation}
where $\mathbf{n}_c$ represents measurement noise, and $\mathbf{M}$ is the binary sampling operator corresponding to the chosen undersampling pattern. 

\subsection{Alternating Training of Sampling and Reconstruction Networks}
In this section, we introduce an alternating training framework that jointly optimizes the sampling network, MNet~\cite{huang2022single}, and a reconstruction network. 
The sampling network, denoted as $\mathcal{M}_{\boldsymbol{\Gamma}}$ with parameters $\boldsymbol{\Gamma}$, predicts the sampling pattern from low-frequency $k$-space, while the reconstruction network $f_\theta$, parameterized by $\theta$, recovers the underlying image from undersampled multi-coil $k$-space.
The alternating updates for training the sampler and reconstructor are thus given by:

\begin{align}
\underset{\boldsymbol{\Gamma}}{\min} \; & 
\sum_{i}\, \varphi \!\big(\mathcal{M}_{\boldsymbol{\Gamma}}(\mathbf{z}_i), \mathbf{M}_i\big)
+ \lambda \frac{\| f_\theta (\mathbf{A}_i^H \hat{\mathbf{M}}_i \mathbf{y}_i) - \mathbf{x}_i \|_2}{\| \mathbf{x}_i \|_2},
\label{eq:mnet_loss} \\
\underset{\theta}{\min} \; & 
\sum_{i} \frac{\| f_\theta (\mathbf{A}_i^H \hat{\mathbf{M}}_i \mathbf{y}_i) - \mathbf{x}_i \|_2}{\| \mathbf{x}_i \|_2},
\label{eq:unet_update}
\end{align}

where $\mathbf{y}_i$ and $\mathbf{x}_i$ are the fully-sampled multi-coil $k$-space and corresponding ground-truth image for the $i$th training scan, $\mathbf{z}_i$ is the low-frequency $k$-space averaged across coils, and $\mathbf{A}_i = \mathbf{F}\mathbf{S}_i$ denotes the fully sampled multi-coil MRI encoding operator with $\mathbf{A}_i^H$ being its adjoint operator.
$\mathbf{M}_i$ is the reference mask pre-generated using the ICD sampling optimization algorithm~\cite{gautam2026scan}. The ICD sampling algorithm updates individual sampling locations to reduce reconstruction error while preserving the sampling budget. The scalar $\lambda>0$ weights the reconstruction-based regularization term against the mask cross-entropy loss. 

The function $\varphi$ is the binary cross-entropy loss between the MNet-predicted logits
$\mathbf{a}_i=\mathcal{M}_{\boldsymbol{\Gamma}}(\mathbf{z}_i)$ and the corresponding
ICD-generated binary reference mask $\mathbf{M}_i$:
\begin{multline}
    \varphi(\mathbf{a}_i,\mathbf{M}_i)
    =
    -\frac{1}{N_{\mathrm{PE}}}
    \sum_{j=1}^{N_{\mathrm{PE}}}
    \Big[
    M_{i,j}\log \sigma(a_{i,j})\\
    +(1-M_{i,j}) \log\big(1-\sigma(a_{i,j})\big)
    \Big].
\end{multline}
Here, $a_{i,j}$ and $M_{i,j}$ denote the corresponding entries for the $j$th candidate phase-encode line, $N_{\mathrm{PE}}$ denotes the number of candidate phase-encode lines, and $\sigma(\cdot)$ denotes the sigmoid function.

The predicted logits $\mathcal{M}_{\boldsymbol{\Gamma}}(\mathbf{z}_i)$ are used to form the binary mask $\hat{\mathbf{M}}_i$ by setting the top-$B$ entries to one and the remaining candidate entries to zero, according to the sampling budget, and then combining them with the fixed central lines.
Since this selection step is non-differentiable, we use a straight-through
estimator~\cite{bengio2013estimating}, with the pre-selection sigmoid
probabilities used to approximate gradients during backpropagation.

\subsection{Reconstruction Network Training}

The reconstruction network $f_\theta$ is trained to recover the ground truth image $\mathbf{x}_i$ from undersampled multi-coil $k$-space, given a collection of scan-adaptive MNet predicted masks $\hat{\mathbf{M}}_i$. In our framework, we use two different reconstruction networks: UNet~\cite{ronneberger2015u} and MoDL~\cite{aggarwal2018modl}.
For the sampler-reconstructor update in Eq.~\eqref{eq:unet_update}, we use UNet as the reconstruction model $f_\theta$, which takes as input the image obtained by applying the adjoint of the MRI forward operator to the undersampled multi-coil $k$-space, i.e., $\mathbf{A}_i^H \hat{\mathbf{M}}_i \mathbf{y}_i$, and is trained to recover the corresponding ground-truth image.


In addition, we post-train an unrolled MoDL reconstructor~\cite{aggarwal2018modl} on the MNet predicted masks after the sampler-reconstructor update is complete. MoDL combines a learned denoiser with a model-based data consistency (DC) step. MoDL is formulated to optimize
\begin{equation}
\mathbf{x}_{\mathrm{rec}} = \arg \min_{\mathbf{x}} \;\big\| \hat{\mathbf{M}}\mathbf{A}\mathbf{x} - \mathbf{y} \big\|_2^2 
\;+\; \mu \,\big\| \mathbf{x} - \mathcal{D}_\theta(\mathbf{x}) \big\|_2^2,
\end{equation}
where $\mathcal{D}_\theta$ denotes a learned denoiser and the scalar $\mu>0$ balances data consistency against the denoiser prior. This optimization is solved using variable splitting with an auxiliary variable $\mathbf{z}$, leading to the following alternating updates:
\begin{align}
\mathbf{x}_{n+1} &= \arg\min_{\mathbf{x}} \;\big\| \hat{\mathbf{M}} \mathbf{A} \mathbf{x} - \mathbf{y} \big\|_2^2 
+ \mu \,\big\| \mathbf{x} - \mathbf{z}_n \big\|_2^2, \\
\mathbf{z}_{n+1} &= \mathcal{D}_\theta\!\left(\mathbf{x}_{n+1}\right).
\end{align}

The $\mathbf{x}$-update is typically solved with conjugate gradients (CG) ensuring data consistency with the acquired $k$-space, while the $\mathbf{z}$-update applies the learned denoiser $\mathcal{D}_\theta$. After unrolling $K$ such iterations, the final output $\mathbf{x}_{K}$ is used as the MoDL reconstruction.




%% file: writeup/experiments.tex
To evaluate the effectiveness of the proposed framework, we conducted experiments on the publicly available fastMRI multi-coil knee dataset~\cite{zbontar2018fastmri}.

\begin{figure}[ht]
  \centering
  \includegraphics[width=\linewidth]{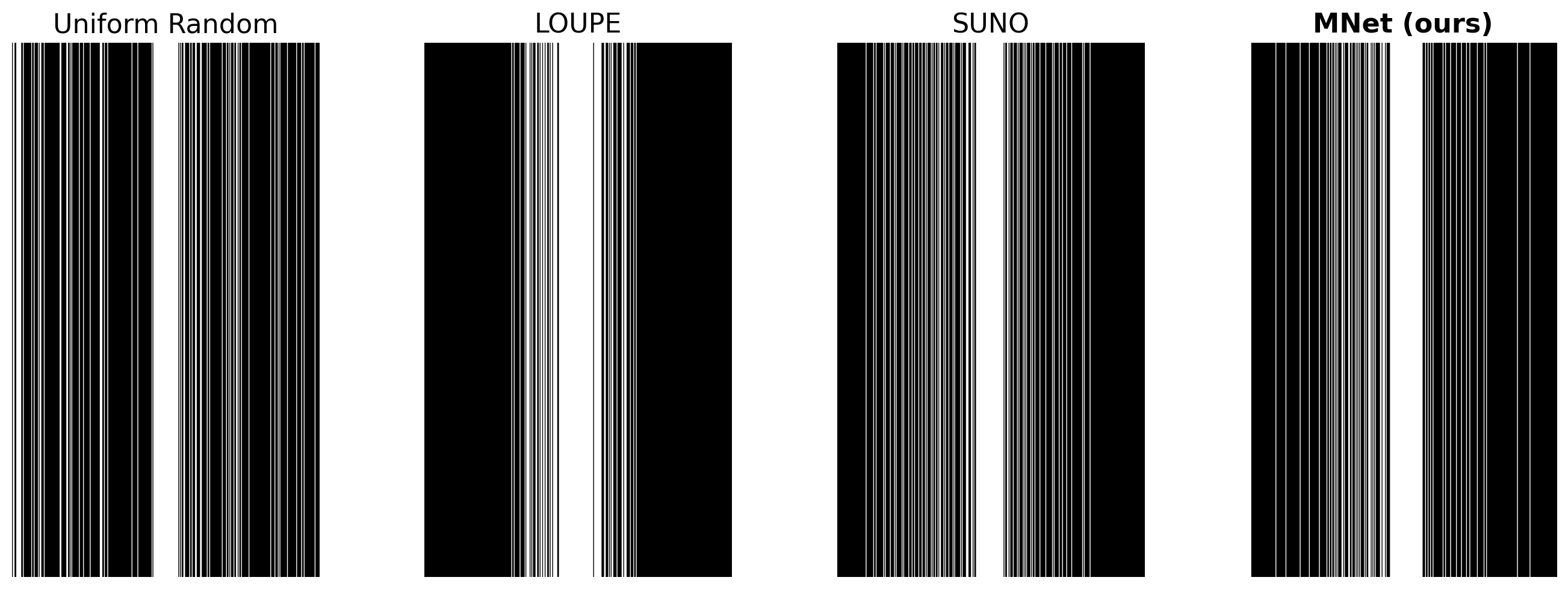}
  \caption{Comparison of undersampling masks at $4\times$ acceleration: (a) uniform random, (b) LOUPE, (c) SUNO and (d) MNet-predicted masks.}
  \label{fig:baselines}
\end{figure}

\begin{figure*}[!t]
  \centering
  \begin{subfigure}[b]{\textwidth}
    \centering
    \includegraphics[width=0.75\textwidth]{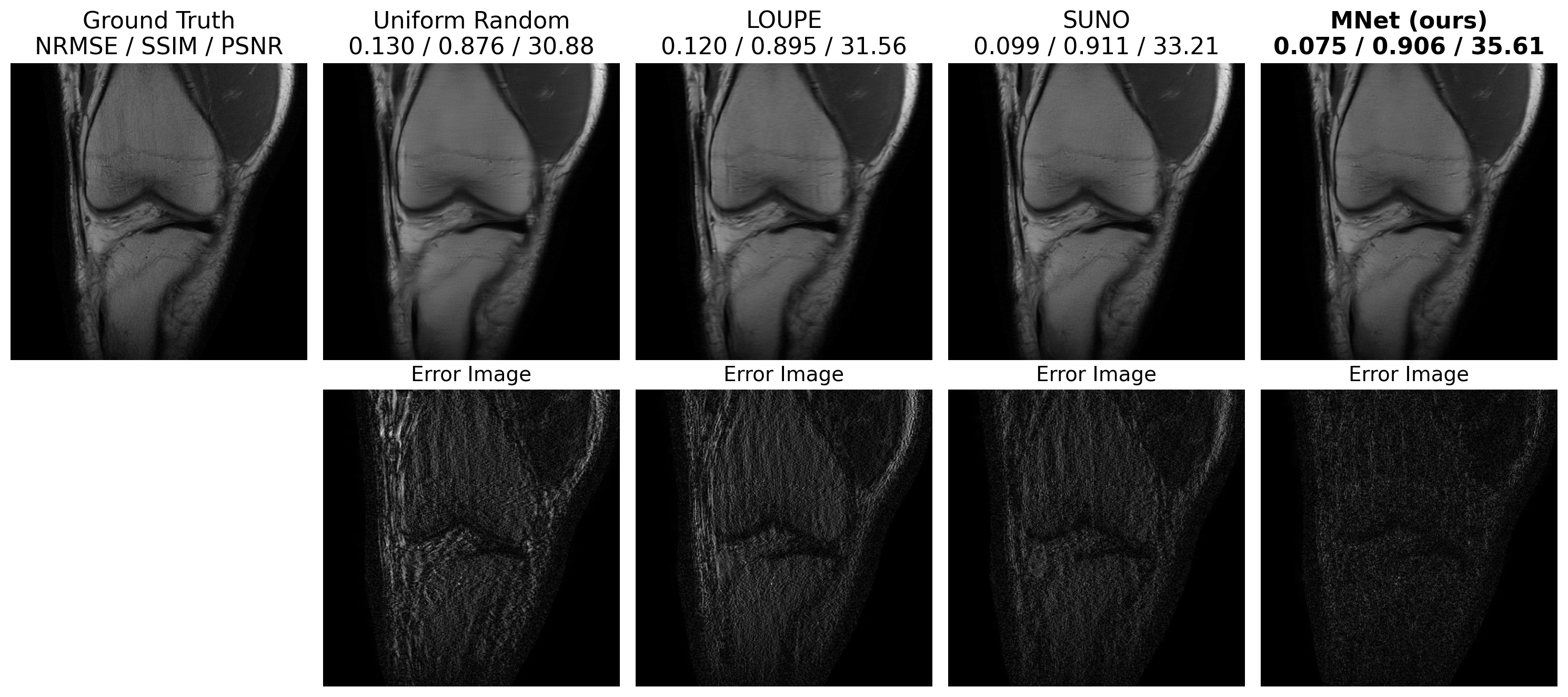}
    \caption{$4\times$ acceleration}
    \label{fig:recons_4x}
  \end{subfigure}

  \vspace{0.6em}

  \begin{subfigure}[b]{\textwidth}
    \centering\includegraphics[width=0.75\textwidth]{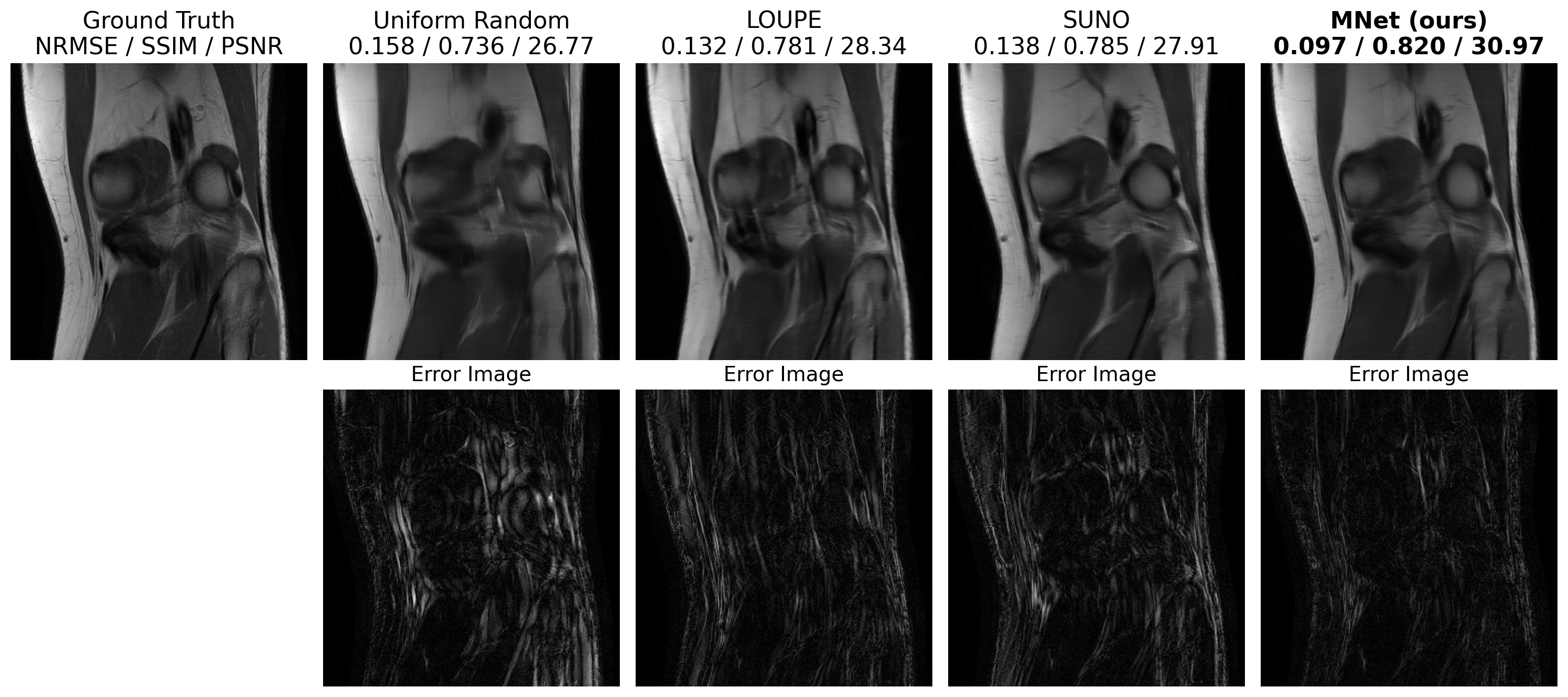}
    \caption{$8\times$ acceleration}
    \label{fig:recons_8x}
  \end{subfigure}
  \caption{MoDL reconstructions at (a) $4\times$ and (b) $8\times$ acceleration for uniform random, LOUPE, SUNO, and the proposed MNet sampling patterns, with corresponding error maps. NRMSE, SSIM, and PSNR (dB) are reported for each reconstruction. MNet-MoDL yields the lowest NRMSE and highest PSNR in both examples, while preserving anatomical structure with reduced reconstruction error.}
  \label{fig:recons1}
\end{figure*}

\subsection{Dataset}
We used the multi-coil fastMRI knee dataset~\cite{zbontar2018fastmri} with measurements from 15 receiver coils and k-space dimensions of 640 × 368. Coil sensitivity maps were estimated from low-frequency $k$-space data using the ESPIRiT algorithm~\cite{pruessmann1999sense, uecker2014espirit}. 
We used 1,438 training slices and 194 testing slices.
The images were cropped to the central $320 \times 320$ region for training the reconstruction network, since it contains most of the useful anatomical content.

\subsection{Baselines}
We compared our method against three baselines: (i) uniform random sampling with fixed low-frequency lines, a widely used non-learning sampling technique, (ii) LOUPE~\cite{bahadir2020deep}, a population-adaptive deep learning method that jointly optimizes sampling patterns and a trained reconstructor, and (iii) SUNO~\cite{gautam2026scan}, the scan-adaptive method that optimizes per-scan masks offline via iterative coordinate descent (ICD) and selects masks at test time through nearest-neighbor search based on acquired low-frequency $k$-space. 
We conducted experiments at $4\times$ and $8\times$ acceleration factors. For each acceleration factor, one-third of the total sampling budget was allocated to fixed central low-frequency lines. Figure~\ref{fig:baselines} shows the MNet-predicted mask for a test scan alongside the uniform random, LOUPE, and estimated SUNO masks at $4\times$ acceleration. 

\begin{table*}[t]
\centering
\setlength{\tabcolsep}{7pt}
\renewcommand{\arraystretch}{1.08}
\footnotesize
\begin{tabular}{ll|ccc|ccc}
\hline
\rule{0pt}{2.4ex}
\textbf{Reconstructor} & \textbf{Sampler}
& \multicolumn{3}{c|}{$4\times$}
& \multicolumn{3}{c}{$8\times$} \\
\cline{3-8}
\rule{0pt}{2.6ex}
& &
\textbf{NRMSE} $\downarrow$ &
\textbf{SSIM} $\uparrow$ &
\textbf{PSNR (dB)} $\uparrow$ &
\textbf{NRMSE} $\downarrow$ &
\textbf{SSIM} $\uparrow$ &
\textbf{PSNR (dB)} $\uparrow$ \\
\hline

\multirow{5}{*}{UNet}
& Uniform Random
& 0.160 $\pm$ 0.063 & 0.749 $\pm$ 0.056 & 28.27 $\pm$ 2.15
& 0.292 $\pm$ 0.123 & 0.622 $\pm$ 0.056 & 23.15 $\pm$ 1.87 \\

& LOUPE
& 0.150 $\pm$ 0.075 & 0.770 $\pm$ 0.055 & 29.13 $\pm$ 2.61
& 0.206 $\pm$ 0.120 & 0.700 $\pm$ 0.076 & 26.59 $\pm$ 3.33 \\

& SUNO
& 0.133 $\pm$ 0.053 & \textbf{0.799 $\pm$ 0.058} & \textbf{29.98 $\pm$ 2.12}
& 0.209 $\pm$ 0.078 & 0.673 $\pm$ 0.073 & 25.95 $\pm$ 2.51 \\

& \textbf{MNet (fixed UNet) (Ours)}
& 0.144 $\pm$ 0.039 & 0.788 $\pm$ 0.054 & 28.92 $\pm$ 2.29
& 0.216 $\pm$ 0.089 & 0.665 $\pm$ 0.056 & 25.64 $\pm$ 2.03 \\

& \textbf{MNet Alternate (Ours)}
& \textbf{0.131 $\pm$ 0.039} & 0.791 $\pm$ 0.053 & 29.86 $\pm$ 1.96
& \textbf{0.179 $\pm$ 0.074} & \textbf{0.716 $\pm$ 0.071} & \textbf{27.33 $\pm$ 2.46} \\

\hline

\multirow{5}{*}{MoDL}
& Uniform Random
& 0.124 $\pm$ 0.039 & 0.820 $\pm$ 0.047 & 30.33 $\pm$ 1.85
& 0.222 $\pm$ 0.073 & 0.715 $\pm$ 0.055 & 25.35 $\pm$ 1.94 \\

& LOUPE
& 0.108 $\pm$ 0.040 & 0.848 $\pm$ 0.047 & 31.69 $\pm$ 2.06
& 0.175 $\pm$ 0.069 & 0.767 $\pm$ 0.057 & 27.49 $\pm$ 2.23 \\

& SUNO
& 0.096 $\pm$ 0.032 & \textbf{0.860 $\pm$ 0.045} & 32.69 $\pm$ 1.63
& 0.162 $\pm$ 0.056 & 0.763 $\pm$ 0.055 & 28.06 $\pm$ 2.04 \\

& \textbf{MNet (fixed UNet) (Ours)}
& 0.100 $\pm$ 0.042 & 0.858 $\pm$ 0.047 & 32.44 $\pm$ 1.85
& 0.157 $\pm$ 0.068 & 0.770 $\pm$ 0.059 & 28.51 $\pm$ 2.34 \\

& \textbf{MNet Alternate (Ours)}
& \textbf{0.095 $\pm$ 0.034} & 0.858 $\pm$ 0.049 & \textbf{32.79 $\pm$ 1.82}
& \textbf{0.152 $\pm$ 0.061} & \textbf{0.777 $\pm$ 0.058} & \textbf{28.73 $\pm$ 2.17} \\

\hline
\end{tabular}
\caption{Reconstruction performance on the fastMRI multi-coil knee test set at $4\times$ and $8\times$ acceleration for different sampling strategies and UNet/MoDL reconstructors. Values are reported as mean $\pm$ standard deviation over the test set. MNet-MoDL achieves the best NRMSE and PSNR at $4\times$, with SSIM comparable to SUNO-MoDL, and achieves the best performance across all three metrics at $8\times$. Best values for each reconstructor and acceleration factor are shown in bold.}
\label{tab:results1}
\end{table*}

\subsection{Training Configuration}
We used two training setups to study the impact of joint optimization: MNet training with an alternately updated UNet and MNet training with a fixed UNet. In the alternating setting, the mask generation network $\mathcal{M}_{\boldsymbol{\Gamma}}$ and reconstruction network $f_\theta$ are updated alternately.
In the MNet training with a fixed UNet, the reconstruction network $f_\theta$ is pretrained on ICD masks and kept fixed, while only the sampler parameters $\Gamma$ are optimized.
The sampler objective also includes the reconstruction error term in Eq.~\eqref{eq:mnet_loss}, which encourages the predicted masks to produce accurate image reconstruction. 
The UNet and MoDL reconstruction models were trained separately for 4$\times$ and 8$\times$ acceleration factors.  A separate MoDL reconstructor was then post-trained for each sampler using its predicted masks on the training set.

During training, MNet was optimized with RMSprop~\cite{graves2013generating} using a learning rate of $10^{-4}$, while the UNet and MoDL reconstruction models were trained with Adam~\cite{kingma2014adam} using a learning rate of $10^{-3}$ for 100 epochs. All the training and inference experiments were performed on an NVIDIA RTX A5000 GPU.
The networks used two-channel (real and imaginary) images as the input and the output. To quantify image reconstruction quality, we used normalized root mean squared error (NRMSE), structural similarity index measure (SSIM), and peak signal-to-noise ratio (PSNR) as metrics.




\subsection{Results}
In this section, we present the quantitative and qualitative reconstruction results for our proposed sampling method along with the baselines at $4\times$ and $8\times$ acceleration.
Figure~\ref{fig:recons1} shows the MoDL reconstructed images for different sampling methods, together with the corresponding error maps and reconstruction metrics at $4\times$ and $8\times$ acceleration. Compared to uniform random and LOUPE sampling, the MNet-based sampling produces sharper reconstructions with reduced residual aliasing and better preservation of fine anatomical structures. These visual improvements are also reflected in the error maps, which show lower reconstruction error for MNet, particularly at $8\times$ acceleration. Consistent with these observations, MNet-MoDL achieves the lowest NRMSE and highest PSNR among the methods shown in both representative examples, while maintaining comparable or improved SSIM.

Quantitative performance across the test set is summarized in Table~\ref{tab:results1}. When the MNet masks obtained from the alternating
MNet-UNet framework are paired with the post-trained MoDL reconstructor, the resulting reconstructions achieve the lowest NRMSE and highest PSNR at
both acceleration factors. At $4\times$, the SSIM is comparable to SUNO, while at $8\times$, MNet Alternate achieves the best performance across all
three metrics. With UNet reconstruction, MNet Alternate similarly achieves performance comparable to SUNO at $4\times$ and the best performance across
all three metrics at $8\times$. Thus, the advantage of the proposed sampling approach becomes more evident at the higher acceleration factor.

The alternating training configuration also consistently improves over MNet trained with a fixed UNet. This difference is relatively small at
$4\times$ but becomes more pronounced at $8\times$, particularly with UNet reconstruction, supporting the benefit of adapting the sampler and reconstructor together under more aggressive undersampling.

\subsection{Mask Prediction Time Comparison}
To assess the computational complexity of our proposed MNet approach, we compared its mask prediction/inference time with that of SUNO. 
For MNet, we measured the inference time required to predict a binary sampling mask from the acquired low-frequency $k$-space measurements.
For SUNO, we measured the time taken for nearest-neighbor search between the reconstruction from test $k$-space and all the training reconstructions~\cite{gautam2026scan}. To evaluate the effect of increasing dictionary size beyond the original 1,438 training slices, we synthetically enlarged the SUNO dictionary by generating additional transformed copies of the training images using rotations and flips, resulting in dictionaries of $N=2{,}000$ and $N=5{,}000$ slices. Unlike SUNO, whose search cost increases with the dictionary size $N$, MNet requires a single forward pass and therefore has prediction cost independent of $N$. As shown in Table~\ref{tab:timing}, the SUNO prediction time increases from 11.5 ms at $N=1{,}438$ to 39.9 ms at $N=5{,}000$, while MNet remains at 10.5 ms. Consequently, the speedup provided by MNet increases from $1.1\times$ to $3.8\times$ over this range. These results demonstrate that MNet provides increasingly faster mask prediction as the SUNO dictionary grows, since its inference cost is independent of the dictionary size. 

\begin{table}[!t]
\centering
\setlength{\tabcolsep}{6pt}
\renewcommand{\arraystretch}{1.15}
\footnotesize
\begin{tabular}{c|c|c}
\hline
\textbf{Dictionary size $N$ (slices)} &
\textbf{SUNO (ms)} &
\textbf{Speedup over MNet} \\
\hline
1,438 & 11.5 & 1.1$\times$ \\
2,000 & 16.0 & 1.5$\times$ \\
5,000 & 39.9 & 3.8$\times$ \\
\hline
\end{tabular}
\caption{SUNO mask prediction time for different dictionary sizes. MNet requires 10.5 ms per mask prediction independent of $N$.}
\label{tab:timing}
\end{table}

%% file: writeup/conclusion.tex


We proposed a scan-adaptive MRI undersampling framework that predicts sampling patterns directly from low-frequency multi-coil $k$-space using a learned CNN-based sampling network. The sampler is trained using pre-optimized ICD masks as supervised labels and is jointly trained with a reconstruction network in an alternating manner. Experiments on the fastMRI multi-coil knee dataset show that the proposed approach achieves better reconstruction quality comparable to baselines at $4\times$ and 8$\times$ acceleration. The alternating MNet configuration achieves the best mean performance across the reported metrics for both UNet and post-trained MoDL reconstruction. The alternating training strategy also consistently improves over training MNet with a fixed UNet, with a larger benefit under more aggressive undersampling. 
 Beyond reconstruction quality, MNet predicts scan-specific masks in constant time independent of the training dictionary size, eliminating the retrieval cost requirements of nearest-neighbor approaches. In the future, we plan to extend the sampling network with a mixture-of-experts design to further enhance its ability to adapt to diverse anatomical structures and scanning conditions.